\documentclass[aps,pre,preprint,showpacs,unsortedaddress,
superscriptaddress]{revtex4}
\usepackage{epsfig}
\usepackage{latexsym}
\begin{document}
\title{Phase separation of binary fluids with dynamic  temperature}
\author{G. Gonnella}
 \affiliation{Dipartimento di
Fisica, Universit\`{a} di Bari,
 {\it and} INFN, Sezione di Bari,
Via Amendola 173, 70126 Bari, Italy}
\author{A. Lamura}
\affiliation{
Istituto Applicazioni Calcolo, CNR,
Via Amendola 122/D, 70126 Bari, Italy}
\author{A. Piscitelli}
\affiliation{Dipartimento di
Fisica, Universit\`{a} di Bari,
 {\it and} INFN, Sezione di Bari,
Via Amendola 173, 70126 Bari, Italy}
\author{A. Tiribocchi}
\affiliation{Dipartimento di
Fisica, Universit\`{a} di Bari,
 {\it and} INFN, Sezione di Bari,
Via Amendola 173, 70126 Bari, Italy}
\date{\today}
\begin{abstract}
Phase separation of binary fluids quenched by contact with cold
external walls is considered. 
Navier-Stokes, convection-diffusion, and energy
equations are solved by lattice Boltzmann method
coupled with finite-difference schemes.
At high viscosity, different morphologies are observed by varying the thermal
diffusivity. In the range of thermal diffusivities with domains growing
parallel to the walls, temperature and phase separation fronts
propagate towards the inner of  the system   with power-law behavior.
At low viscosity hydrodynamics favors rounded shapes, and complex patterns
with  different lengthscales appear. Off-symmetrical systems behave
similarly but with more ordered configurations.
\end{abstract}
\maketitle
\section{Introduction}

When in a multi-phase system initially in a mixed state the
temperature is decreased
to values corresponding to
a coexisting region of the phase diagram,
domains of ordered
phases start to form and grow with time. The process is called phase
separation and is relevant for a large variety of
systems \cite{GUNTON}. In most of the cases studied theoretically,
the temperature or other
control parameters are assumed  not depending on time and
space, but are instantaneously set to their final values for coexistence.
This assumption, reasonable  in many situations,  typically gives
rise to a self-similar growth behavior with a characteristic domain
size following a time power-law \cite{BRAY}. However, there are cases where
the dynamics  of
 the control parameter needs to be considered \cite{SAARLOS}
since it can
  greatly  affect the morphology of domains.
In binary alloys, for example, slow cooling is  used to
produce optimal sequences of alternate bands of different materials
\cite{METAL}.
In
polymeric mixtures the possibility of controlling the demixing
morphology
 by appropriate thermal driving has been studied in
Refs.~\cite{KREKHOV,YAMAMURA};
modulated patterns have been observed when a mixture is periodically
brought
above and below the critical point
\cite{TANAKA}. Other worth examples of
 complex pattern formation  due to  the dynamics of the
control parameters occur
 in  crystal growth  \cite{LANGER},  immersion-precipitation membranes
 \cite{CHENG}, or in electrolyte diffusion in gels  \cite{LIESEGANG,ANTAL}.

In this paper we study binary fluids
 quenched by contact with cold walls
at temperatures below the critical value.
 The behavior of binary fluids
 in  sudden  quenches at homogeneous  temperature is quite known
\cite{BRAY,YEOMANS}. For  symmetric composition,  the typical
interconnected pattern of spinodal decomposition is observed. In the
system here considered, phase separation is expected to start close
to the walls and develop in the inner of the system following the
temperature evolution. The dynamics of this process and the role of
the velocity field have not been explored too much, in spite of
their relevance for many of the systems mentioned above.

Two-dimensional  studies  of diffusive binary systems with cold
sharp fronts propagating at constant speed have shown the formation
of structures aligned on a direction depending on the speed
\cite{FURUKAWA,ANTAL,HANTZ,WAGNERFOARD,KREK2}. These results
are also supported by theoretical analysis
\cite{WAGNERFOARD,KREK2}. Lamellar-like  structures have
been also found in numerical studies of two-dimensional
off-symmetrical binary systems with  the temperature following a
fixed diffusive law \cite{BALL}. In a model with the temperature
dynamically coupled to the concentration field, point-like cold
sources have been shown to give rise to ring structures of alternate
phases \cite{ESPANOL}. On the other hand, more usual morphologies
have been found in cases with fixed thermal gradient \cite{JASNOW},
while complex phenomena such as sequential phase-separation cascades
have been observed when the control parameter is slowly
homogeneously changed \cite{VOLLMER}.
 The effects of full coupling
between all thermo-hydrodynamic variables have been not considered
sofar.

The paper is organized as follows. In the next section the theoretical model
and the numerical method
are illustrated. The dynamics of our system is described by
mass, momentum, and energy equations with thermodynamics based
on a free-energy functional including gradient terms.
In Section III the results of our simulations
are shown. 
We will explore the control parameter space by varying the viscosity
and the 
thermal diffusivity. This will allow to analyze the differences with respect
to the behavior of binary fluids in instantaneous quenching.
The presentation will be focused on few cases typical for each regime.
A final discussion will follow in Section IV.

\section{The model}

We consider a binary mixture
with  dynamical variables $T,${\bf v}$, n, \varphi$ which
are,  respectively, the temperature, the velocity, the total density,
and the order parameter field being the concentration difference.
Equilibrium
properties are encoded in the free-energy
\begin{equation}
F=\int
(\psi(n,\varphi,T)+\frac{1}{2}M|{\bf \nabla}
\varphi|^{2})d{\bf r}
\end{equation}
where
\begin{equation}
\psi(n,\varphi,T)= e -
k_{B}T[n\ln(n)-\frac{n+\varphi}{2}\ln(\frac{n+\varphi}{2})-
\frac{n-\varphi}{2}\ln(\frac{n-\varphi}{2})]
\end{equation}
with $ e = n k_B T +
\frac{\lambda n}{4}(1-\frac{\varphi^{2}}{n^{2}})$ 
being the bulk internal energy 
and the term
in square brackets the mixing entropy.
The gradient term in Eq.~(1) is a combination 
of an internal energy gradient contribution proportional
to $K$ and of an entropic term proportional to 
$-C$ \cite{ONUKI}, hence $M=K+CT$. 
The system has a critical transition at $k_B T_c =
\lambda/2$ and the  order parameter in the separated phases takes
the values $\varphi_{\pm}(T) = \pm \sqrt{3n^2(T_c/T -1)}$.
 The  dynamical equations are  given by \cite{DGM}
\begin{equation}
\partial_t n=-\partial_{\alpha} (n v_{\alpha}), \label{mass}
\end{equation}
\vskip -0.4cm
\begin{equation}
\partial_t \varphi=-\partial_{\alpha} (\varphi
v_{\alpha})-2\partial_{\alpha} J_{\alpha}^d, \label{massdiff}
\end{equation}
\vskip -0.4cm
\begin{equation}
\partial_t (n v_{\beta}) = - \partial_{\alpha} (n v_{\alpha} v_{\beta})
-\partial_{\alpha}(\Pi_{\alpha \beta}-\sigma_{\alpha \beta}),
\label{momentum}
\end{equation}
\vskip -0.4cm
\begin{equation}
\partial_t \widehat{e}=-\partial_{\alpha}(\widehat{e}
v_{\alpha})-(\Pi_{\alpha \beta}-\sigma_{\alpha \beta})\partial_{\alpha}
v_{\beta}-\partial_{\alpha}J^{q}_{\alpha} \label{ener},
\end{equation}
where ${\bf J}^{d}$ and ${\bf J}^{q}$ are the
diffusion and heat currents, $\Pi_{\alpha \beta}$ is the reversible stress
tensor, $\sigma_{\alpha \beta}=\eta(\partial_{\alpha} v_{\beta} +
\partial_{\beta} v_{\alpha}) +(\zeta-2 \eta/d)\delta_{\alpha \beta}
\partial_{\gamma} v_{\gamma}$ is the
dissipative stress tensor with  $\zeta, \eta$ being  the bulk and
shear  viscosities, respectively, $d$ the space dimension, and
$\widehat{{e}}=e +  \frac{K}{2}|{\bf \nabla}\varphi|^{2} $ the
total internal energy density  also including gradient
contributions. We have recently established the expressions for the
pressure tensor $\Pi_{\alpha \beta}$ and chemical potential $\mu$
\cite{GONN} following the approach of Ref.~\cite{ONUKI}. One finds
\begin{equation}
\Pi_{\alpha \beta}=\left(p - M\varphi
\nabla^{2}\varphi - M{|{\bf \nabla}\varphi|^{2}}/2 -
T\varphi{\bf \nabla}\varphi\cdot{\bf \nabla}
({M}/{T})\right)\delta_{\alpha \beta} +
M\partial_{\alpha}\varphi\partial_{\beta}\varphi
\end{equation}
  where $p = - \psi +
n\partial \psi/\partial n + \varphi
\partial \psi/\partial \varphi$ and
$\mu=\partial \psi/\partial \varphi|_T - T {\bf \nabla} \cdot
[(M/T){\bf \nabla}\varphi]$.
 Finally, in order to
completely set up the dynamical system,
phenomenological expressions for the currents are needed.
As usually, one takes
${\bf J}^{d}=-\mathcal{L}_{11}
{\bf \nabla}({\mu}/{T})+\mathcal{L}_{12}{\bf \nabla}
({1}/{T})$, 
${\bf J}^{q}=-\mathcal{L}_{21}{\bf \nabla}({\mu}/{T})
+\mathcal{L}_{22} {\bf \nabla}({1}/{T}) $ where
$\mathcal{L}_{\alpha \beta}$ is the positively defined matrix of kinetic
coefficients with $\mathcal{L}_{11} = T \Gamma$ and $\mathcal{L}_{22}
= T^2 k$, $\Gamma$ and $k$ being the mobility and thermal diffusivity, 
respectively, assumed constant \cite{DGM}.

In order to solve
Eqs.~(\ref{mass}-\ref{ener}) in $d=2$ we have developed a hybrid
lattice Boltzmann method (LBM) \cite{lall,xu,maren,STELLA} 
where LBM \cite{LBM} is
used to simulate the continuity and Navier-Stokes equations
(\ref{mass}) and (\ref{momentum}) while finite-difference methods are
implemented to solve the convection-diffusion
 and the energy equations (\ref{massdiff}) and (\ref{ener}). LBM
 has been widely
used to study multi-phase/component fluids \cite{DUN} and, in particular,
hydrodynamic effects in phase ordering \cite{CATES}. It is defined
in terms of a set of distribution functions, $f_i({\bf r},t)$ with
$i=0,1,...,8$, located in each site ${\bf r}$ at each time $t$ of a
D2Q9 (2 space dimensions and 9 lattice velocities)
lattice where sites are connected to first and second neighbors
by lattice velocity vectors of modulus $|{\bf e}_i|=c$ 
($i=1,...,4$) and $|{\bf e}_i|=\sqrt{2}c$ ($i=5,...,8$),
respectively. The zero velocity vector ${\bf e}_0=0$ is also
included. The lattice speed is $c=\Delta x/\Delta t$ where $\Delta
x$ and $\Delta t$ are the lattice and time steps, respectively. The
distribution functions evolve according to a single relaxation time
Boltzmann equation \cite{bgk} supplemented by a forcing term
\cite{guo}
\begin{equation}\label{evoleqn}
f_i({\bf r}+{\bf e}_i\Delta t,t+\Delta t)-f_i({\bf r},t)=-\frac{\Delta t}
{\tau}[f_i({\bf r},t)-f_i^{eq}({\bf r},t)]+\Delta t F_i({\bf r},t),
\end{equation}
where $\tau$ is the relaxation parameter, $f_i^{eq}$ are the
equilibrium distribution functions, and $F_i$ are the forcing terms
to be properly determined.

The total density and the fluid momentum are given by the following
relations
\begin{equation}\label{moment}
n=\sum_if_i , \hspace{1.3cm} n{\bf v}=\sum_if_i{\bf e}_i
+ \frac{1}{2}{\bf F}\Delta t,
\end{equation}
where ${\bf F}$ is the force density acting on the fluid. The
$f_i^{eq}$ are expressed   as a standard second order expansion in
the fluid velocity ${\bf v}$ of the Maxwell-Boltzmann distribution
functions \cite{qian}.
The forcing terms $F_i$ in Eq.~(\ref{evoleqn})
are expressed as a second order expansion
in the lattice velocity vectors \cite{LADD}.
The continuity and the Navier-Stokes equations
(\ref{mass}) and (\ref{momentum}) can be recovered  by using a
Chapman-Enskog expansion
when the $F_i$ are given by
\begin{equation}\label{latticeforceterm}
F_i=\left(1-\frac{\Delta t}{2\tau}\right)\omega_i\left[
\frac{{\bf e}_i-{\bf v}}{c^2_s}+\frac{{\bf e}_i\cdot{\bf v}}{c^4_s}
{\bf e}_i\right]\cdot {\bf F}
\end{equation}
with the force density ${\bf F}$  having components
\begin{equation}
F_{\alpha}=\partial_{\alpha}(nc_s^2)-\partial_{\beta}\Pi_{\alpha
\beta} ,
\end{equation}
$c_s=c/\sqrt{3}$ being the speed of sound in the LBM,
$\omega_0=4/9$, $\omega_i=1/9$ for $i=1,...,4$, and $\omega_i=1/36$
for $i=5,...,8$.
We observe that in this formulation  the pressure tensor is inserted
as a body force in the lattice Boltzmann equations. From the
Chapman-Enskog  expansion  it comes out that $\xi=\eta$ with 
\begin{equation}
\eta = n c_s^2 \Delta t\left(\frac{\tau}{\Delta t}
- \frac{1}{2}\right).
\end{equation}

On the other hand, a two-step finite difference scheme is used for
the equations (\ref{massdiff}) and (\ref{ener}) (details on the
implementation of Eq.~(\ref{massdiff}) in the case of an isothermal
LBM can be found in Ref.~\cite{STELLA}). At walls, no-slip boundary
conditions are adopted for the LBM \cite{PHYSA}, the temperature is
set to fixed values 
$T_b$ at the bottom wall and $T_u$ at the up wall, respectively, 
and neutral wetting for the concentration is adopted. 
This latter condition corresponds to impose
${\bf a} \cdot \nabla \varphi|_{walls}= 0$
and ${\bf a} \cdot \nabla (\nabla^2 \varphi)|_{walls}=0$, where ${\bf a}$ is an
inward normal unit vector to the walls. These conditions together ensure  
${\bf a} \cdot \nabla \mu|_{walls}= 0$ so that the concentration
gradient is parallel to the walls and there is
no flux across the walls.
We have found this algorithm stable in a wide range of
temperatures, viscosities and thermal diffusivities. With
respect to thermal LBM for non-ideal fluids \cite{SOFO}
where lattice Boltzmann equations are used
to simulate the full set of macroscopic  dynamical equations, the
present model allows to reduce the number of lattice velocities thus
speeding up the code and reducing the required memory \cite{STELLA}. 

\section{Results and discussion}

In the following we will explore the parameter space keeping fixed the values 
of $K=0.003, C=0, k_B T_c = 0.005,
\Gamma=0.1$, and $\mathcal{L}_{12}=\mathcal{L}_{21}=0$. We will use lattices
of size ranging from $256 \times 256$ to $1024 \times 1024$.
We have considered different values of $\eta$ and $k$.
Before focusing on the cases representative of the various regimes,
we will list all the runs we did in terms of dimensionless numbers. 

Common numbers used in hydrodynamics are the Reynolds
and Peclet numbers $Re$ and $Pe$. 
They are defined as $Re=v L/\nu$, where $\nu=\eta/n$ is the kinematic 
viscosity,  $Pe_{md}=v L/D$ for mass diffusion, where $D$ is the mass 
diffusion coefficient, and $Pe_{td}=v L/k$ 
for thermal diffusion. 
$L$ and $v$ are a typical length and velocity of the system.
In phase separation $L$ can be identified with the average size
of domains so that $Re$ and $Pe$ would depend on time (for a discussion see
Ref.~\cite{kendon}). It is therefore more convenient for our purposes to 
introduce the Schmidt  and Prandtl  numbers $Sc$ and $Pr$
defined as $Sc=\nu/D$ and $Pr=\nu/k$, where
$D= |a| \Gamma$  
with $a=(k_B T_c/n) (T/T_c-1)$ being the coefficient of the linear term 
in the chemical potential $\mu$ \cite{KREKHOV,GONN}. Here 
$T$ can be chosen as the
value of the temperature at the walls.
Table I contains a list of the runs we did, 
reported in terms of $Sc$ and $Pr$.
It is also useful to evaluate the Mach number 
$Ma=|{\bf v}|_{max}/c_s$ where
$|{\bf v}|_{max}$ is the maximum value of the fluid velocity
during evolution.
In all our simulations $Ma$ is always much smaller than $0.1$ 
(see in the following), and
the fluid results practically incompressible, as checked, 
with $n \simeq 1$. For this reason we do not present in the paper
any result about the time evolution of the total density $n$.

First, as a benchmark for  our
 method, we consider the  relaxation of a single
interface profile with $k=10^{-2}$ and $\eta=0.167$ ($\tau=1$). 
This corresponds to a low viscosity regime as discussed in the following.
We started the simulation with a sharp concentration step with values
$\varphi_{-}(T_b)$ and $\varphi_{+}(T_u)$ and bulk temperature
$T/T_c=0.8$ keeping fixed the temperatures $T_b/T_c=0.8,
T_u/T_c=0.9$  at the bottom and up walls (Fig.~1 (a)). The system
reaches a stationary state with constant temperature gradient and
concentration profile as in Fig.~1 (b). The numerical values of
concentrations in the two bulk phases are in  very good agreement
with the analytical expression for $\varphi_{\pm}(T({\bf r}))$ 
corresponding to the equilibrium values of $T({\bf r})$ 
shown in the related inset. This means that the concentration
field $\varphi$ is in local equilibrium. The
temperature of the up wall is then set to the same value of the
temperature of the bottom wall (Fig.~1 (c)).
 Then, as it can be seen in Fig.~1 (d), the system
equilibrates at constant temperature with the expected concentration
profile. Spurious velocities
are of order $10^{-9}$
 and result completely negligible. The test shows
 that stationary  states are well reproduced   by our algorithm.

\subsection{Diffusive regime}

We describe our results for phase separation. We  first
consider a case at very high viscosity with $\eta=6.5$ ($\tau=20$)
and symmetric composition
(Runs 1-8). Here
the effects of the velocity field are negligible. We set
$T_b/T_c=T_u/T_c=0.8$ and initial bulk temperature above $T_c$. As
it can be seen in Fig.~2,  for thermal diffusivities $k \ge 10^{-1}$,
usual isotropic phase separation is observed.
 In the  range
$ k= 5 \times 10^{-4} \div 5 \times 10^{-2}$, in spite of the
neutral wetting condition  on the boundaries, domains   in the bulk
have interfaces preferentially  parallel  to thermal fronts.
 For smaller values of $k$  domains grow
perpendicularly to the walls. These results agree  with those of
Refs.~\cite{FURUKAWA,WAGNERFOARD,KREK2} 
in purely diffusive models
where the same morphological
sequence was found by   decreasing   the speed of cold fronts moving
into a region with the mixed phase. However, also in absence of
hydrodynamic effects,  our case is different since the
thermodynamics of the mixture is fully consistently treated and
temperature fronts have no sharp imposed profile.

We will now concentrate on cases at intermediate thermal diffusivities
where domains are parallel to the walls
and propagation fronts can be traced.
Concentration and temperature  configurations at successive times
for $k=10^{-2}$ (Run 4a) are shown in  Fig.~3 and  Fig.~4, respectively.
In this case it is $Ma \simeq 5 \times 10^{-5}$.
The temperature fronts have typical diffusive profiles which slowly
relax   to the equilibrium value imposed on the boundaries. In
order to be quantitative, we defined $y_T(t)$ as the distance from
the wall where the temperature assumes a fixed value (we chose
$T/T_c=0.88$) and measured this quantity in simulations with   large
rectangular lattices. 
The solution of the diffusion
equation with initial  temperature $T_0$ and fixed boundary value
$T_w$ is $(T(y,t)-T_w)/(T_0-T_w)=erf{[y/(2 \sqrt{k t})]}$ 
which implies   $y_T/\sqrt{k} \sim \sqrt{t}$.
In the inset of Fig.~5 it is shown, in simulations with  different $k$, that
$y_T$ follows the  standard diffusion behavior.
The time behavior of $y_T$ has been checked not depending on the specific
value of the ratio $T/T_c$ in the range $[0.8,1.0]$; by considering 
a value of $T$ such that $T/T_c<1$ allows to track the position of the
temperature front for a longer time interval.

One can also consider the behavior of the fronts limiting the
regions with separated phases, clearly
observable in the first three snapshots of Fig.~3.
 Their position can be defined  as the distance $y_{\varphi}$
from the walls beyond which the condition
$\nabla \varphi \simeq 0$ is verified everywhere. 
More precisely, we took $y_{\varphi}$ as the point beyond which
$|\nabla \varphi| < C$ with $C=\sqrt{2} \times 0.01$; the value of $C$
is chosen to match the maximum value of the fluctuations 
of $|\nabla \varphi|$ in the initial
disordered state, where $|\varphi| < 0.01$.
(In the
last snapshot of Fig.~3 the two fronts propagating from up and down
have come close each other and more usual phase separation occurs in
the central region of the system.) We measured $y_{\varphi}$ on
rectangular lattices for different $k$ and  observed
 deviations
  from
diffusive behavior
(see Fig.~5). We found that $y_{\varphi}$ grows by power law
with an exponent depending on $k$. Our fits give $y_{\varphi} \sim
t^{0.66}$ for $k=10^{-2}$ and exponents closer to $1/2$ for smaller
$k$. 
We analyzed for different $k$ possible variations 
of the typical values of fluid velocity but we did not find any. 
Therefore
the change of the exponent of $y_{\varphi}$ cannot be attributed
to the velocity field. 
Even if $y_{\varphi}$ moves faster than $y_T$ and at long times it results
$y_{\varphi} > y_T$, we checked that the relation $y_{\varphi} < y_{T_c}$
is always verified so that phase separation always occurs for $T < T_c$. 
Since the phase separation is induced by the
temperature change, one could
have expected a similar behavior for   $y_{\varphi}$ and $y_T$.
The discrepancy  could be related to the broad character
of the temperature fronts which spreads  the phase separated region.
We also observed that the width of lamellar domains decreases at
larger $k$, in agreement with Ref.~\cite{WAGNERFOARD}.

\subsection{Hydrodynamic regime}

At lower  viscosities the evolution of morphology is very different
in   the range with  intermediate values of thermal diffusivity. 
We will in particular illustrate in Fig.~6 the case with 
 $\eta=0.167$ ($\tau=1$) and $k=10^{-2}$ (Runs 19),
for which we found  $Ma \simeq 5 \times 10^{-4}$.
This is the same thermal diffusivity of Fig.~3.
At this viscosity hydrodynamics is relevant. 
Indeed, in
instantaneous quenching at constant temperature and $\eta=0.167$ we
observed the domain growth exponent to 
assume the inertial value $2/3$ (at odd with the diffusive
high-viscosity value $1/3$ ) \cite{YEOMANS}.
The growth exponent was calculated by measuring the  characteristic length
defined by the inverse of the first momentum of the structure
factor \cite{corb}.
The main effect due to hydrodynamics  observable  in Fig.~6 is 
that domains do not grow aligned with
temperature fronts as it occurs for the same thermal diffusivity
at high viscosity.
Circular patterns  are
stabilized by the flow \cite{YEOMANS} and an example is given in
Fig.~7. 
A similar  picture
occurs for other values of $k$ here not reported (see Table I).
On the other hand, the other thermal diffusivity regimes are less
affected by hydrodynamics.
When decreasing $k$, it is still possible to observe 
domains growing with interfaces normal to the walls
as in the case at high viscosity (see Fig.~8 - Run 21b),
while at larger $k$ (Run 18) phase separation occurs isotropically
like in an instantaneous quenching.

The cases shown in Figs.~3 and 6 
are typical of the high and low viscosity regimes.
At intermediate values of $\eta$ one can observe features common to the
two above cases (see Fig.~9 for $\eta=2.167$ - Run 11a).
Concerning the behavior of $y_T(t)$, we could not find relevant differences
by varying $\eta$ with respect to the case at
high viscosity.

Another effect induced by hydrodynamics is the formation 
of structures in the inner part of the system
at earlier times than in the case at high
viscosity (compare Fig.~3 and Fig.~6). In the inner region we can observe
the typical interconnected pattern of spinodal decomposition but
with a characteristic length-scale different from that of domains
close to the walls. However, while the  structures close to the
walls are
in local equilibrium, that is $\varphi({\bf r})=\varphi_{\pm}(T({\bf
r}))$, in the middle of the system the concentration field is such
that $|\varphi| < \varphi_{+}(T({\bf r}))$. A temporal regime
characterized by the presence of  domains with two scales  was found  in
systems of different size (from $256 \times 256$ to $1024 \times
1024$) and $k=10^{-3} \div 10^{-2}$. 
In order to characterize the two scales we analyzed the behavior of the
structure factor. In Fig.~10 the spherically averaged structure factor
is shown at two consecutive times for a system having the same
parameters of Fig.~6 and size $L=512$. Two peaks are observable at each time
that can be interpreted as related to
the existence of two different length scales with one
about twice longer than the other. The higher peak at smaller wave vector 
corresponds to the larger domains close to the walls while
the other peak is related to the thinner domains in the inner of the system.
At increasing times, the two peaks
tend to merge.
Due to this
morphological evolution, in simulations at low viscosity,
 the position of the phase separation
front $y_{\varphi}$ could be measured only for a short time interval
making  not possible to determine the power-law behavior.

Finally, we show results for systems with asymmetric composition. In
Fig.~11 the evolution of two systems only differing  for the value of
viscosity is shown. Lamellar patterns prevail at high viscosity
while circular droplets  dominate at low viscosity ($\eta=0.167$). In
the latter case, again, two typical scales can be observed with thin
tubes of materials connecting larger domains.  The behavior of
 $y_T$ is similar to that of the symmetric case.

\section{Conclusions}

We have developed a numerical method for thermal
binary fluids described by continuity, Navier-Stokes, convection-diffusion, 
and energy  equations. 
We have studied quenching by contact with external
walls, and we have shown how the pattern formation depends on thermal
diffusivity, viscosity, and composition of the system.
The evolution is very different from that observed in instantaneous
homogeneous quenching.
At high
viscosity, different orientations of domains are possible. In an intermediate 
range of thermal diffusivities  domains are parallel to the walls.
The fronts limiting the regions with separated domains 
 move towards the inner of the system 
with   a power law behavior 
not always corresponding to that of the temperature fronts. 
At low viscosity, the velocity field favors more circular
patterns, and domains are characterized by different
length-scales close to the walls and 
in the inner of the system. Off-symmetrical
mixtures give more ordered patterns.

We conclude with two remarks on possible future directions of work.
The first one concerns the Soret effect, which
corresponds to have a mass diffusion current induced by thermal
gradients. This effect can become relevant in quenching very close
to the critical point where the ratio  $D_T/D$ becomes large \cite{KREKHOV}.
Here $D_T$ is the thermal (mass) diffusion coefficient
($D_T=\mathcal{L}_{12}/T^2$ in our notation) and $D$ is the mass
diffusion coefficient  defined  at the beginning of Section III.  
In order  to have a
first idea on how the Soret effect can affect  the pattern
morphology, we considered a case with $D_T/D = 20 $  corresponding to
the highest values for this ratio reported in literature \cite{KREKHOV}. 
This would give  $D_T= 2 \times 10^{-3}$,  
taking  for $D$ the value used in the runs of Section III.
We run simulations for this case.
We observed, in
the intermediate range of thermal diffusivity and at high viscosity,
the tendency of the system to exhibit more ordered lamellar patterns
(parallel to the walls).  At higher thermal diffusivity isotropic
phase separation is found as usually, while at very low thermal
diffusivity ($k=10^{-4}$), parallel patterns are found instead of
perpendicular patterns. At low viscosity (we tested the case
corresponding to that  of  Fig.~6) hydrodynamics continues to favor
domains with more circular shape. We run also simulations with
$D_T=10^{-4}$, corresponding to a ratio $D_T/D \simeq 1$, without
finding relevant differences with the respect to the case with
$D_T=0$. 
We also observe that the behavior of $y_{\varphi}$
could depend on our choice for 
$\mathcal{L}_{12}$ and $\mathcal{L}_{21}$.
A more comprehensive analysis of the Soret effect  will be
presented elsewhere.

Finally, the morphology 
could be still richer  in three dimensions, also due to the existence of more
hydrodynamic regimes \cite{BRAY},
so that three-dimensional simulations would  complete
the picture given sofar.

\begin{acknowledgments}
GG warmly acknowledges discussions with A. J. Wagner
during his visit at North Dakota State University.
\end{acknowledgments}

\newpage

\begin{table}[th]
\begin{center}
\begin{tabular}{|p{2.4cm}|p{2.5cm}|p{2.2cm}|p{2cm}|p{2cm}|}
\hline Run & Size & $Sc$ ($\times 10^3$) & $Pr$ & Symbol\\
\hline 1  & 512  & 65 & 12 & I\\
       2a, 2b & 512, 256 & 65 & 66 & I\\
       3 &  512 & 65 & 129 & Pa \\
       4a, 4b & 512, 256 & 65 & 651 & Pa\\
       5a, 5b & 512, 256 & 65 & 1299 & Pa\\
       6 & 256 & 65 & 6500 & Pa\\
       7 & 512 & 65 & 65000 & Pe\\
       8 & 256 & 65 & 650000 & Pe\\
       9 & 256 & 21.7 & 22 & I\\
       10 & 256 & 21.7 & 43 & I, Pa\\
       11a, 11b & 512, 256 & 21.7 & 217 & I, Pa\\
       12 & 256 & 21.7 & 2167 & Pe \\
       13 & 256 & 21.7 & 21667 & Pe\\
       14 & 256 & 8.3 & 8 & I\\
       15 & 256 & 8.3 & 83 & I*\\
       16 & 256 & 8.3 & 833 & I*, Pe\\
       17 & 256 & 8.3 & 8333 & Pe\\
       18 & 512 & 1.7 & 3 & I\\
       19a, 19b, 19c & 1024, 512, 256 & 1.7 & 17 & I* \\
       20a, 20b & 512, 256 & 1.7 & 167 & I*\\
       21a, 21b & 512, 128 & 1.7 & 1667 & Pe\\
\hline
\end{tabular}
\end{center}
\caption{The first column indexes the simulation run,
the second one is the linear size of the lattice, the third one is 
the Schmidt number ($Sc$), the fourth one is the Prandtl number
($Pr$). The
last column is the symbol that identifies the kind of different observed
patterns: I (isotropic morphology), Pa (domains parallel to the walls), 
Pe (domains perpendicular to the walls), I* (isotropic morphology
with two lengthscales).
The runs with two symbols exhibit patterns with common features to those
corresponding to each symbol.}
\label{table1}
\end{table}

\clearpage

\newpage

\begin{figure}[h]
\epsfig{file=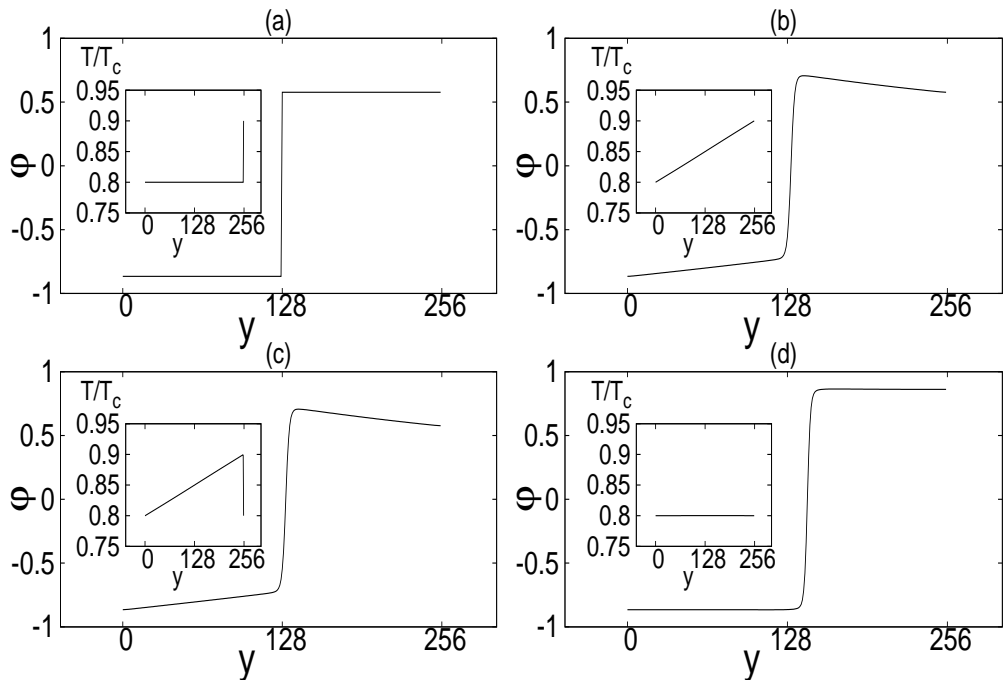,bbllx=-80 pt,bblly= 50
pt,bburx= 710 pt,bbury= 1600 pt,height=14cm,width=9cm,clip=,angle=270}
\caption{Concentration
 and temperature (inset) profiles for an interface
relaxation
(see the text for explanation).
}
\label{fig1}
\end{figure}

\newpage

\begin{figure}[h]
\epsfig{file=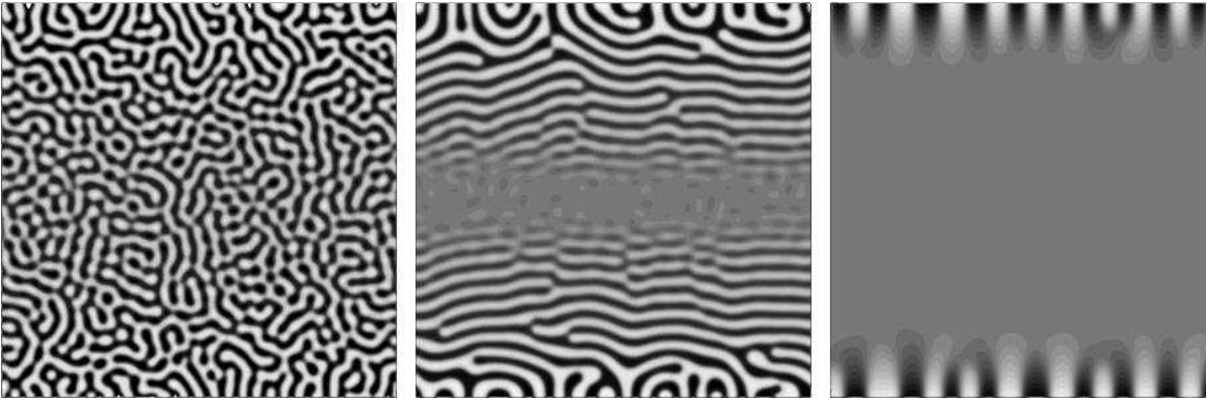,bbllx=0 pt,bblly= 0 pt,bburx= 910 pt,bbury=
320 pt,width=16cm,clip=} \caption{Typical configurations of the
concentration field $\varphi$ for symmetric composition at
very high viscosity ($\eta=6.5$)
with $k=10^{-1}, 10^{-2}, 10^{-5}$ (from left to
right) at times $t=12.5 \times 10^5; 37.5 \times 10^5; 300 \times 10^5$, 
respectively, with lattice size $512 \times 512$, and $T_b/T_c=T_u/T_c=0.8$.}
\label{fig2}
\end{figure}

\newpage

\begin{figure}[h]
\epsfig{file=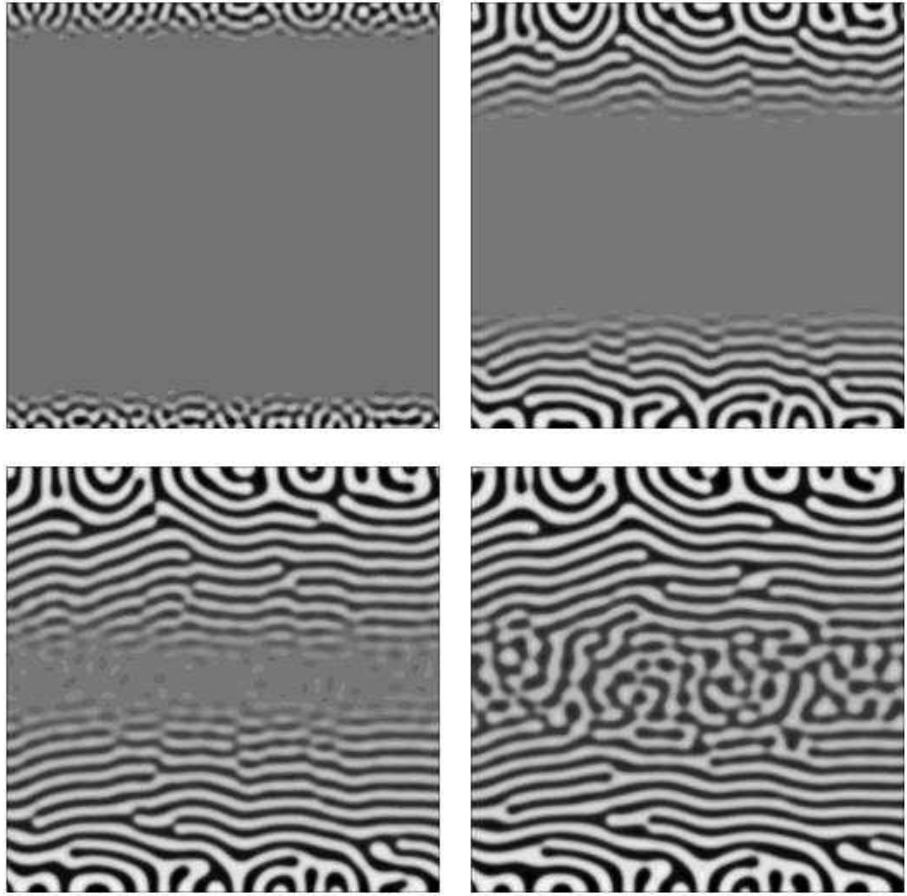,bbllx=40 pt,bblly= 6
pt,bburx= 567 pt,bbury= 648 pt,width=12cm,clip=}
\caption{Configurations of concentration $\varphi$ for composition $50/50$
at times $t=7.5 \times 10^5; 22.5 \times 10^5; 37.5 \times 10^5;
50 \times 10^5$, at very
high viscosity ($\eta=6.5$) with lattice size $512 \times 512$,
$T_b/T_c=T_u/T_c=0.8$, and $k=10^{-2}$.}
\label{fig3}
\end{figure}

\newpage

\begin{figure}[h]
\epsfig{file=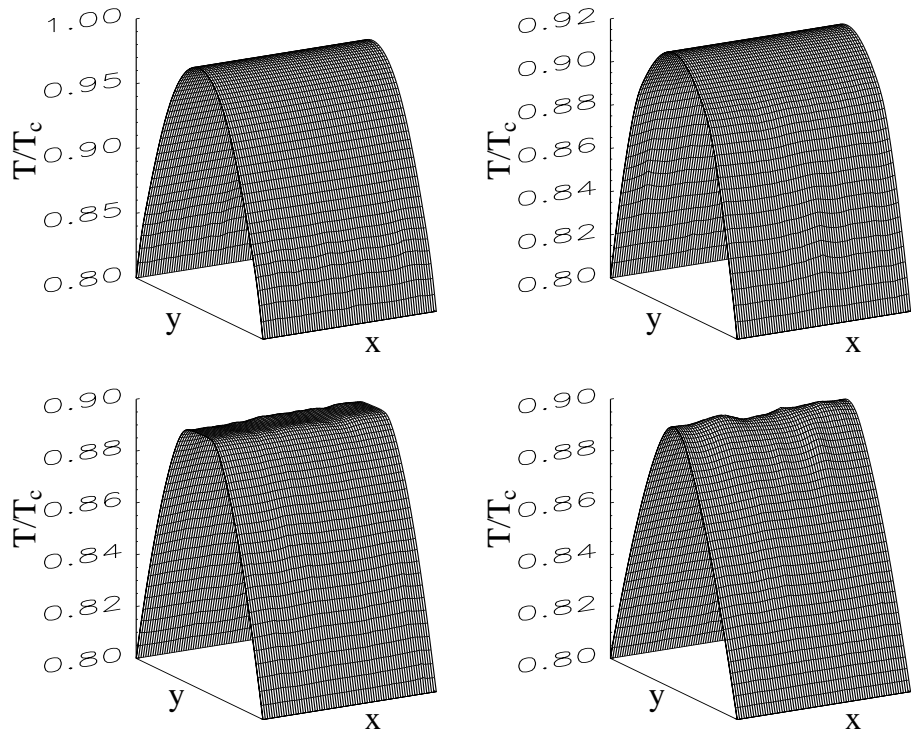,bbllx=00 pt,bblly= 0 pt,bburx= 850 pt,bbury=
860 pt,width=12cm,clip=} \caption{Configurations of the ratio
$T/T_c$ for the
same case and at same times of Fig.~3. Coordinates on the $x$ and $y$
axes are in lattice units and both of them are in the range $[0,512]$.} 
\label{fig4}
\end{figure}
\newpage

\begin{figure}[h]
\epsfig{file=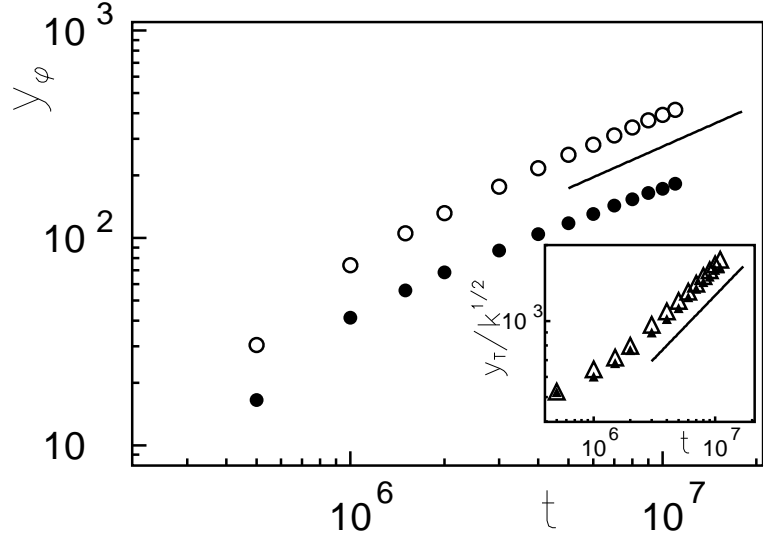,bbllx=73 pt,bblly= 341
pt,bburx= 522 pt,bbury= 700 pt,width=10cm,clip=}
\caption{Time behavior of $y_{\varphi}$ at
$k=10^{-2}$ (empty symbols) and $10^{-3}$ (filled symbols) 
at very high viscosity with lattice size $128 \times 2048$.
The straight
line is a guide to the eye and has slope $2/3$.
Inset: Time behavior of $y_T/\sqrt{k}$
at
$k=10^{-2}$ (empty symbols) and $10^{-3}$ (filled symbols). The straight
line has slope $1/2$.}
\label{fig5}
\end{figure}

\newpage

\begin{figure}[h]
\epsfig{file=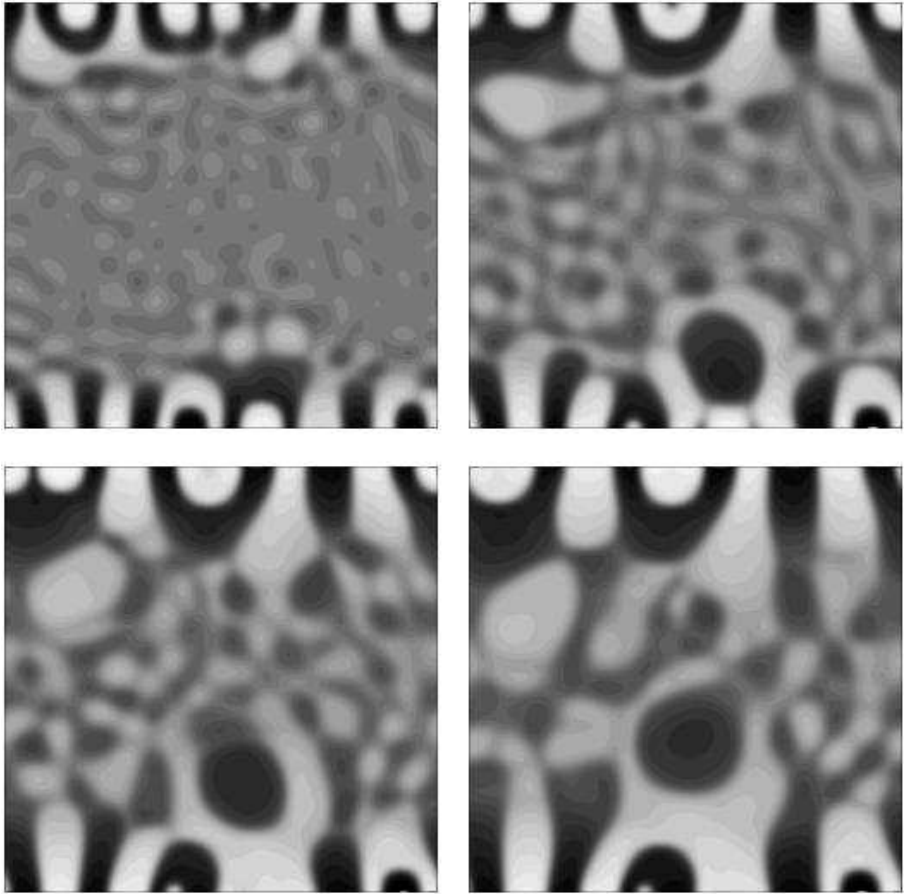,bbllx=40 pt,bblly= 6
pt,bburx= 567 pt,bbury= 648 pt,width=12cm,clip=}
\caption{Configurations of concentration $\varphi$ for composition $50/50$
at times $t=8 \times 10^5 ; 11 \times 10^5; 13 \times 10^5; 15 \times 10^5$,
low viscosity ($\eta=0.167$), lattice size $256 \times 256$,
$T_b/T_c=T_u/T_c=0.8$, and $k=10^{-2}$.}
\label{fig6}
\end{figure}

\newpage

\begin{figure}[h]
\epsfig{file=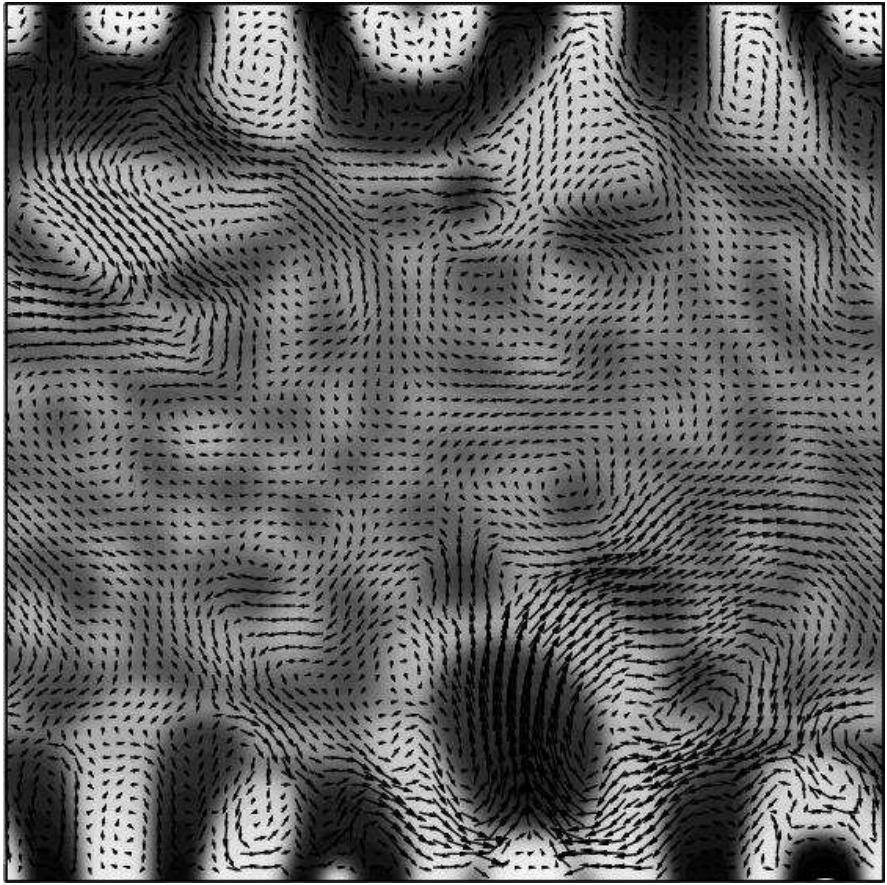,bbllx=0 pt,bblly= 0
pt,bburx= 578 pt,bbury= 788 pt,width=12cm,clip=}
\caption{Configuration of concentration $\varphi$ for 
the case of Fig.~6
at time $t=11 \times 10^5$
with superimposed the velocity field.}
\label{fig7}
\end{figure}

\newpage

\begin{figure}[h]
\epsfig{file=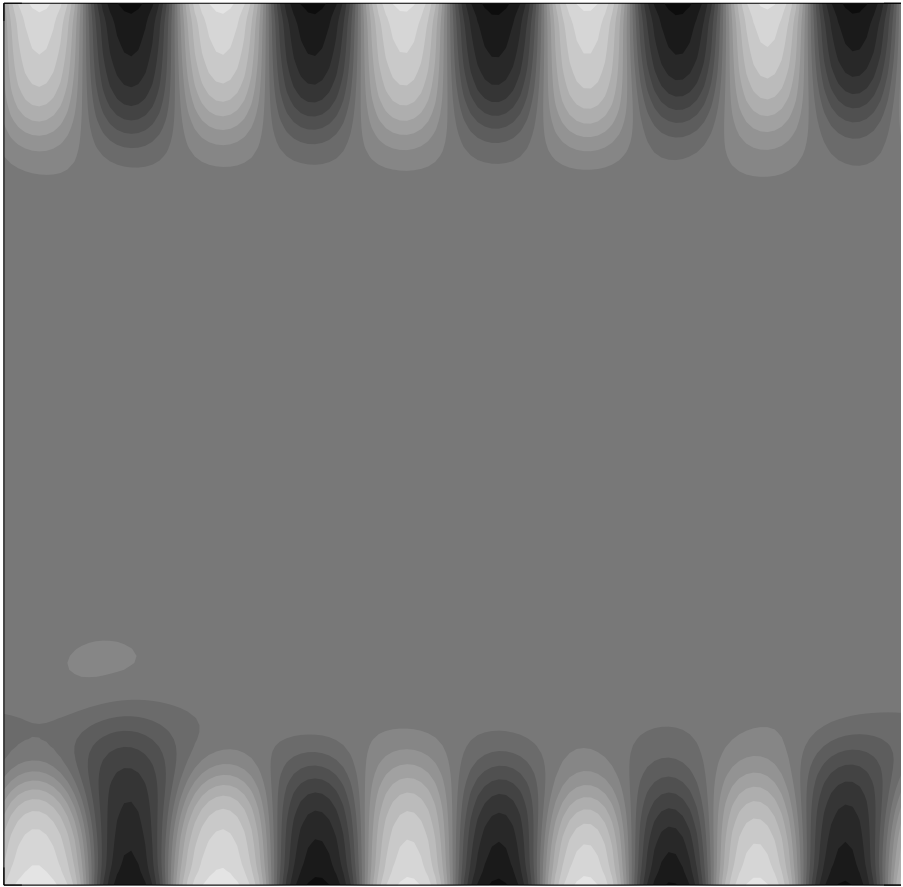,bbllx=115 pt,bblly= 399
pt,bburx= 490 pt,bbury= 769 pt,width=12cm,clip=}
\caption{Configuration of concentration $\varphi$ 
at time $t=14 \times 10^5$, low viscosity ($\eta=0.167$) 
as in Fig.~6, lattice size $128 \times 128$,
$T_b/T_c=T_u/T_c=0.8$, and $k=10^{-4}$.}
\label{fig8}
\end{figure}

\newpage

\begin{figure}[h]
\epsfig{file=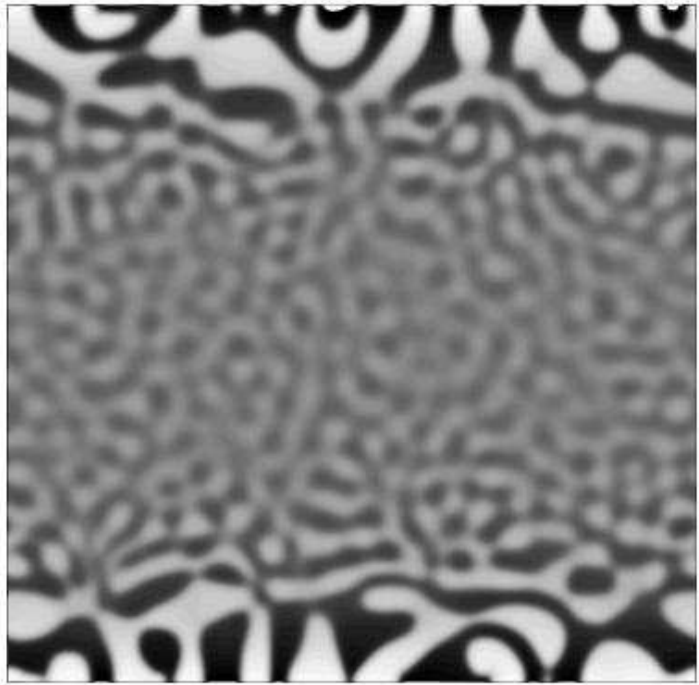,bbllx= 0 pt,bblly= 0
pt,bburx= 490 pt,bbury= 769 pt,width=12cm,clip=}
\caption{Configuration of concentration $\varphi$ 
at time $t=21 \times 10^5$, intermediate viscosity ($\eta=2.167$),
lattice size $512 \times 512$,
$T_b/T_c=T_u/T_c=0.8$, and $k=10^{-2}$.}
\label{fig9}
\end{figure}

\newpage

\begin{figure}[h]
\epsfig{file=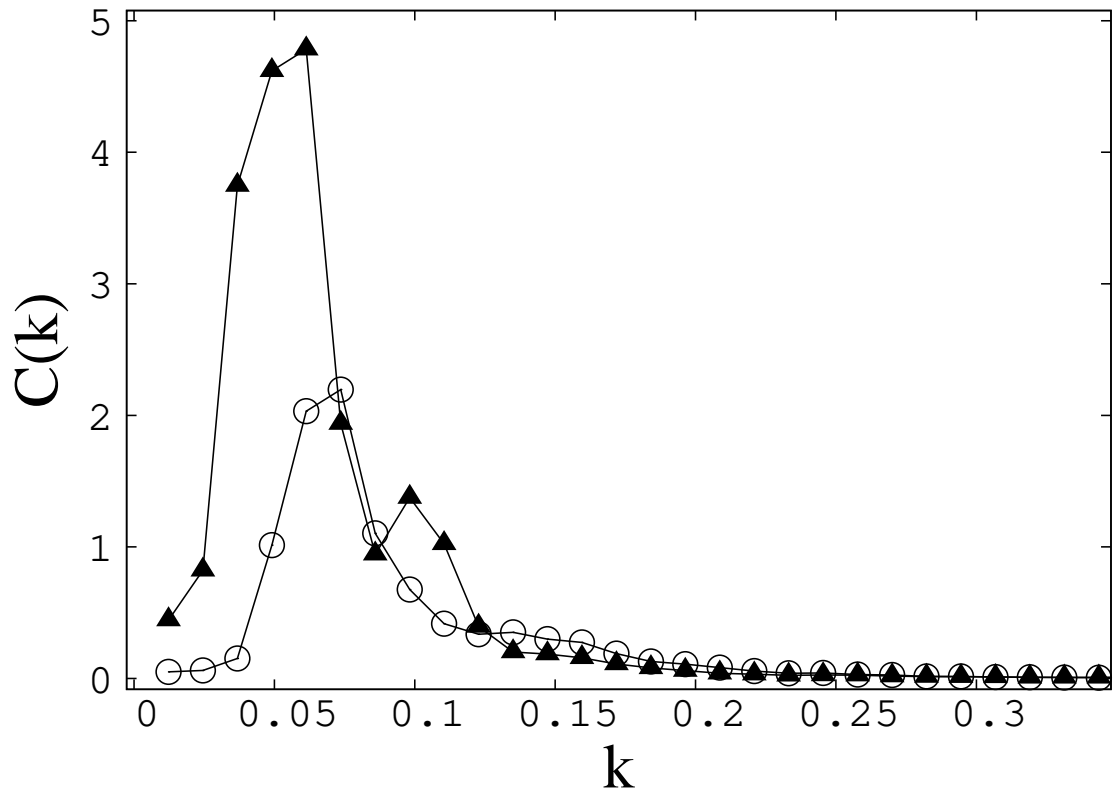,bbllx=15 pt,bblly= 39
pt,bburx= 590 pt,bbury= 769 pt,width=12cm,angle=-90,clip=}
\caption{Spherically averaged structure factor $C(k)$ 
as a function of the wave vector modulus $k$ 
for a system with the same parameters of Fig.~6 and size $L=512$ 
at times $t= 24 \times 10^5$ (empty symbols) and
$t=39 \times 10^5$ (filled symbols), corresponding to the regime with two
scales shown in Fig.~6.
}
\label{fig10}
\end{figure}

\newpage

\begin{figure}[ht]
\epsfig{file=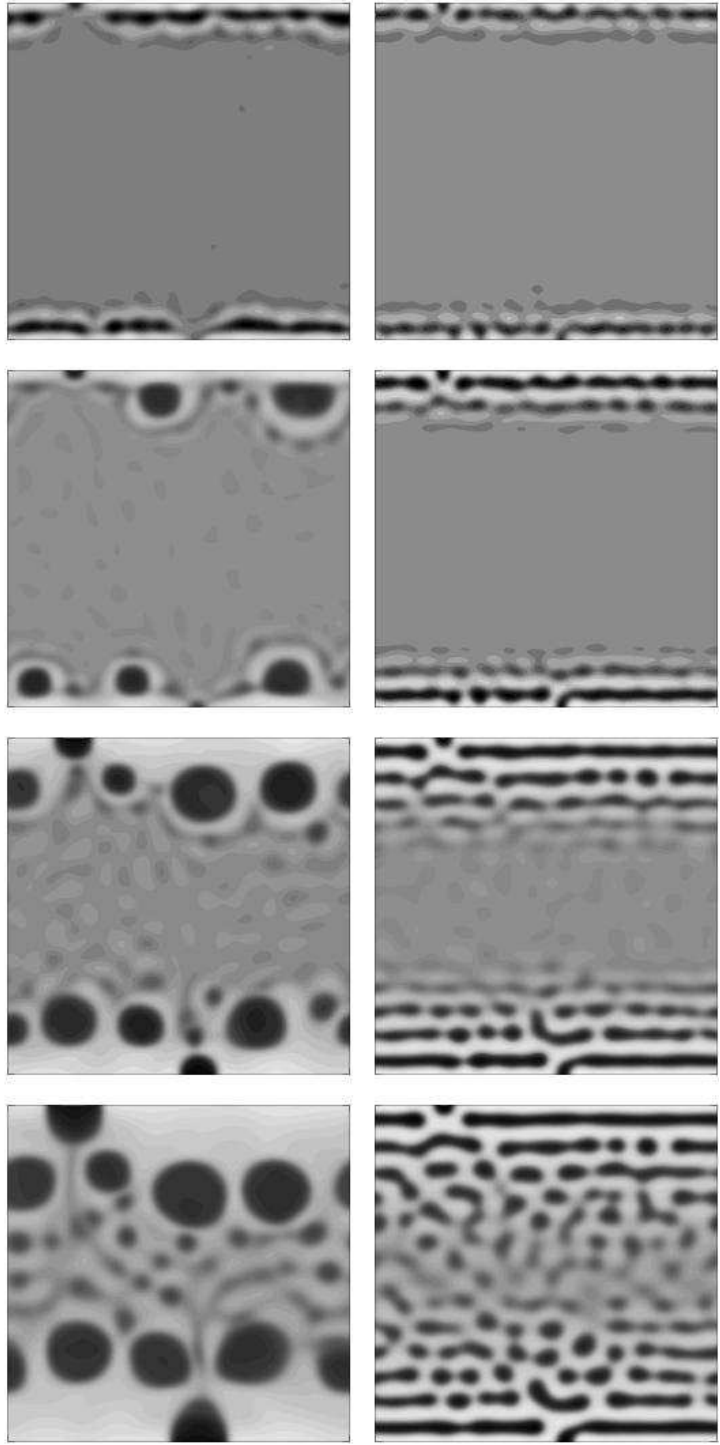,bbllx=40 pt,bblly= 0
pt,bburx= 567 pt,bbury= 1098 pt,width=9.5cm,clip=}
\caption{Configurations of concentration $\varphi$ for composition $55/45$
at times $t=4 \times 10^5 ; 6 \times 10^5; 11 \times 10^5; 16 \times 10^5$,
low viscosity (left column) and very high viscosity (right column),
lattice size $256 \times 256$,
$T_b/T_c=T_u/T_c=0.8$, and $k=10^{-2}$.
Except for the composition, here the parameters are the same used in Figs.~6 
and 3.}
\label{fig11}
\end{figure}

\end{document}